\documentclass[preprint,showpacs,preprintnumbers,amsmath,amssymb]{revtex4}
\usepackage{graphicx}
\usepackage{dcolumn}

\begin{document}

\preprint{APS/123-QED}

\title{Quasifission and fusion-fission in  massive nuclei
reactions. Comparison of reactions leading to the $Z$=120 element
}

\author{A.K. Nasirov}
 \altaffiliation{Institute of Nuclear Physics,Tashkent,
 Uzbekistan}\email{nasirov@jinr.ru}
 \affiliation{Joint Institute for Nuclear Research, Dubna, Russia}
 \author{G. Giardina}%
\affiliation{Istituto Nazionale di Fizica Nucleare, Sezione di Catania, \\
and Dipartimento di Fisica dell'Universit\`a di Messina, 98166, Messina, Italy}%
\author{F. Hanappe}
\affiliation{ Université Libre de Bruxelles, 1050 Bruxelles,
Belgium}%
\author{S. Heinz}
\affiliation{Gesellschaft f\"ur Schwerionenforschung, Darmstadt,
Germany}%
\author{S. Hofmann}
\affiliation{Gesellschaft f\"ur Schwerionenforschung, Darmstadt,
Germany}%
\author{G. Mandaglio}
\affiliation{Istituto Nazionale di Fizica Nucleare, Sezione di Catania, \\
and Dipartimento di Fisica dell'Universit\`a di Messina, 98166, Messina, Italy}%
\author{M. Manganaro}
\affiliation{Istituto Nazionale di Fizica Nucleare, Sezione di Catania, \\
and Dipartimento di Fisica dell'Universit\`a di Messina, 98166, Messina, Italy}%
\author{A.I. Muminov}
\affiliation{Institute of Nuclear Physics, Tashkent, Uzbekistan}%
\author{W. Scheid}
\affiliation{Institut f\"ur Theoretische Physik der
Justus-Liebig-Universit\"at, Giessen, Germany}%

\date{\today}
\begin{abstract}
The yields of evaporation residues, fusion-fission and
quasifission fragments in the $^{48}$Ca+$^{144,154}$Sm and
$^{16}$O+$^{186}$W reactions are analyzed in the framework of the
combined theoretical method based on the dinuclear system concept
and advanced statistical model. The measured yields of evaporation
residues for the $^{48}$Ca+$^{154}$Sm reaction can be well
reproduced. The measured yields of fission fragments are
decomposed into contributions coming from fusion-fission,
quasifission, and fast-fission. The decrease in the measured yield of quasifission
fragments in $^{48}$Ca+$^{154}$Sm at the large collision energies
and the lack of quasifission fragments in the $^{48}$Ca+$^{144}$Sm
reaction are explained by the overlap in mass-angle distributions of
the quasifission and fusion-fission fragments. The investigation
of the optimal conditions for the synthesis of the new element
$Z$=120 ($A$=302) show that the $^{54}$Cr+$^{248}$Cm reaction is
preferable in comparison with the $^{58}$Fe+$^{244}$Pu and
$^{64}$Ni+$^{238}$U reactions because the excitation function of
the evaporation residues of the former reaction is some orders of
magnitude larger than that for the last two reactions.
\end{abstract}

\maketitle

\section{\label{Introduc}Introduction}

The observed evaporation residues in experiments are a result of
the de-excitation of a heated and rotating compound nucleus formed
in complete fusion reactions at heavy ion collisions. There are
no evaporation residues if a compound nucleus is not formed.
The correct estimation of the cross section of the compound nucleus
formation in the reactions with massive nuclei is an important but
difficult task. Different assumptions about the fusion process are
used in different theoretical models and they can give different
cross sections. The experimental methods used to estimate the fusion
probability depend on the unambiguity of identification of the
complete fusion reaction products among the quasifission products. The
difficulties arise when the mass (charge) and angular
distributions of the quasifission and fusion-fission fragments
strongly overlap depending on the reaction dynamics. As a result,
the complete fusion cross sections may be overestimated. We know
that quasifission fragments show anisotropic angular distributions
\cite{Toke85,Shen87}. This is a way to separate them from the
fusion-fission fragments which should have isotropic angular
distributions. But fission fragments in reactions with heavy ions
also show anisotropic angular distributions which is explained by
the assumption that an equilibrium $K$-distribution is not reached
($K$ is the projection of the total spin of the compound nucleus
on its axial symmetry axis). According to the transition state
model \cite{Halpern,Griffin} the formation of a compound nucleus
with a large angular momentum leads to a large anisotropy $A$
which is proportional to $<\ell^2>$:
\begin{equation}
A=1+\frac{\hbar^2<\ell^2>}{4J_{eff}T_{sad}}
\end{equation}
where
\begin{equation}
J_{eff}=J_{\|}J_{\bot}/(J_{\bot}-J_{\|})
\end{equation}
is the effective moment of inertia of the compound nucleus on the
saddle-point; $J_{\|}$ and $J_{\bot}$  are moments of inertia
around the axis of the axial symmetry  and a perpendicular axis,
respectively. $T_{\rm sad}$ is its effective temperature at the
saddle-point. At the same time the angular distribution of the
quasifission fragments may be isotropic when the  dinuclear
system decays having a large angular momentum \cite{Nasirov}.

This paper is devoted to analyze  reasons for the lack or
disappearance of the quasifission feature in the experimental data
for the  $^{48}$Ca+$^{144}$Sm and $^{48}$Ca+$^{154}$Sm reactions
presented in the paper \cite{Knyazheva} by Knyazheva {\it et al.}, as
well as the comparison of the results for these reactions with the
ones for the $^{16}$O+$^{186}$W reaction where there is no
hindrance for complete fusion  \cite{Knyazheva}. The same
method of analysis is applied to study the problem of the synthesis of
the new superheavy element $Z$=120. The three reactions
$^{54}$Cr+$^{248}$Cm, $^{58}$Fe+$^{244}$Pu and $^{64}$Ni+$^{238}$U are
compared with the aim to answer the question which of these
reactions is preferable to obtain $Z$=$^{302}$120.

At first we consider  the $^{48}$Ca+$^{154}$Sm reaction which
shows evidently a yield of quasifission fragments at low
energies. The results in detail were presented in Ref.
\cite{Knyazheva}. According to the conclusion of the authors of
this paper the yield of quasifission fragments disappears by
increasing the beam energy. The model of the dinuclear system
predicts the presence of the quasifission features at large
energies too \cite{GiardinaNP671,Giardina00,faziolett,Nasirov05}.
Another interesting phenomenon is that the authors of Ref.
\cite{Knyazheva} did not observe any yield of quasifission
fragments in the $^{48}$Ca+$^{144}$Sm reaction whereas in the
present work we found a strong hindrance for the complete fusion in
this reaction. The conclusions of the experimental investigation
in \cite{Knyazheva} and our studies are in complete agreement for
the very mass (charge) asymmetric $^{16}$O+$^{186}$W reaction:
the hindrance to complete fusion is negligible. The results of
calculation of the above-mentioned phenomenon are discussed in
Section 2. In Section 3, we present the results of estimating the
evaporation residue yields to find the preferable reaction for the
synthesis of the superheavy element $Z$=120.

\section{\label{overlap}Overlaps of the fusion-fission and
quasifission fragment distributions}

All heavy ion reaction channels with the full momentum transfer at
low collision energies take place through the stage of the
dinuclear system (DNS) formation and can be called capture
reactions. In the deep inelasic collisions DNS is formed but
the full momentum transfer does not occur. Therefore,
the deep inelasic collisions are not capture reactions.
In the capture reactions the colliding nuclei are trapped
into the well of the nucleus-nucleus potential after
dissipation of part of the initial relative kinetic energy
and orbital angular momentum. The lifetime of DNS should be
enough for its transformation into compound nucleus during
its evolution.
The formation of the compound nucleus (CN) in reactions
with massive nuclei has a hindrance: not all of the dinuclear
systems formed at capture of the projectile by the target-nucleus
can be transformed into CN. The decay of the DNS into two
fragments bypassing the stage of the CN formation we call
quasifission. The fast-fission process is the inevitable decay of
the fast rotating mononucleus into two fragments without reaching
the equilibrium compact shape of CN. Such a mononucleus is formed
from the dinuclear system which survived against quasifission. At large
values of the angular momentum $\ell > \ell_f$, where $\ell_f$ is
a value of $\ell$ at which the fission barrier of the
corresponding compound nucleus disappears,
the mononucleus immediately decays into two fragments
 \cite{Gregoire}.  As distinct from
fast-fission, the quasifission can occur at all values of $\ell$
at which capture occurs.

\begin{figure}
\vspace*{-1.0cm}
\begin{center}
\resizebox{0.80\textwidth}{!}{\includegraphics{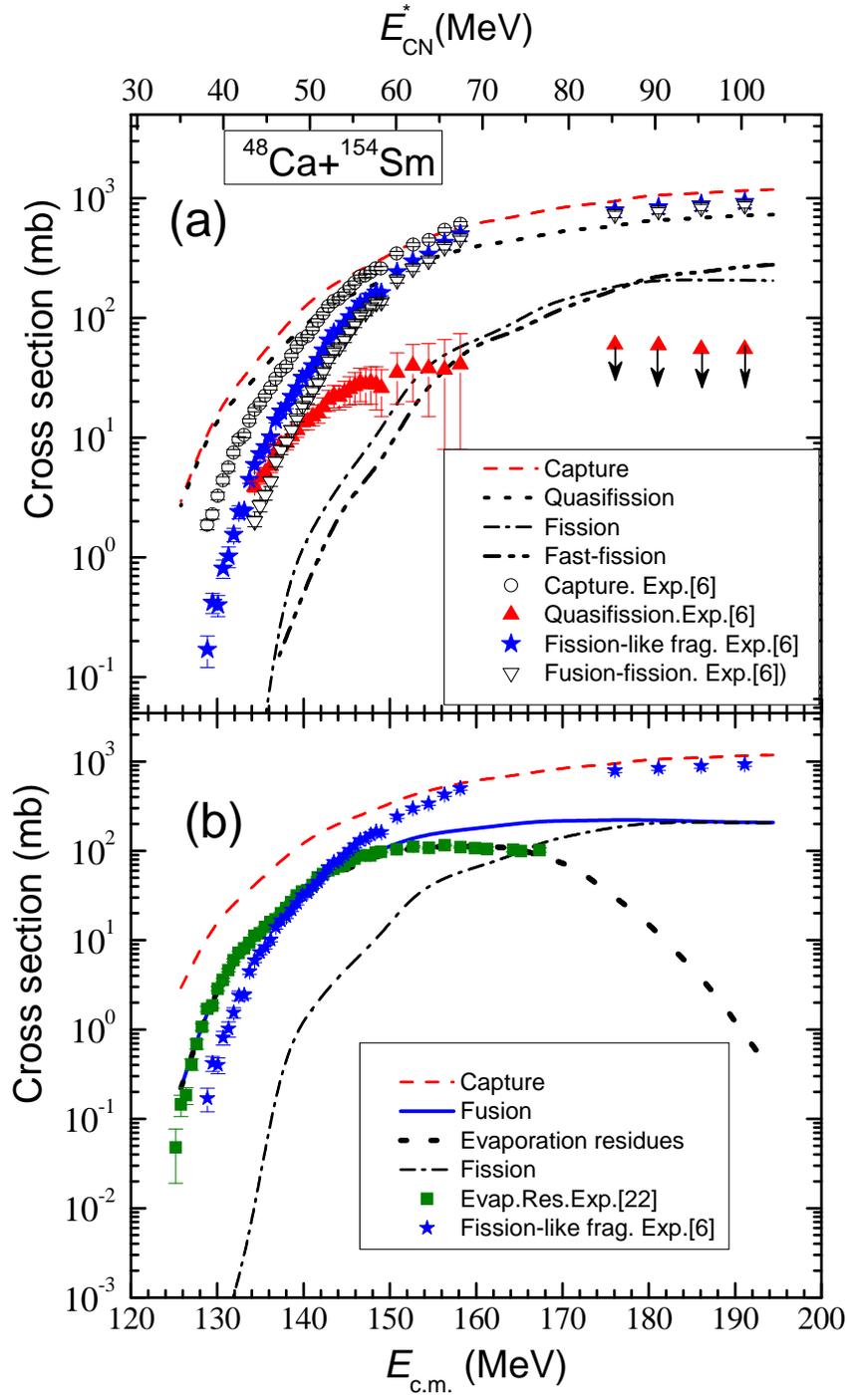}}
\vspace*{-0.5cm} \caption{\label{compcross} Comparison of the
results of this work by the DNS model for the capture, complete
fusion, quasifission, fast-fission and evaporation residue cross
sections with the measured data of the fusion-fission and
quasifission given in Ref. \cite{Knyazheva} (panel (a)) and with
data of the evaporation residues obtained from Ref.
\cite{Stefanini} (panel (b)) for the $^{48}$Ca+$^{154}$Sm
reaction.}
\end{center}
\end{figure}

In Ref. \cite{Knyazheva} the authors established the fusion suppression
and the presence of quasifission for the reactions with the deformed
$^{154}$Sm target at energies near and below the Coulomb barrier.
 But the authors did not analyze the
products with masses outside the range $55<A<145$. In the
mass distribution of fission fragments from the $^{48}$Ca+$^{154}$Sm reaction
they found an ``asymmetric fission mode'' appearing as ``shoulders''
peaked around the masses 65 and 140 at $E^*_{\rm CN}$=49 and 57 MeV.
Quasifission cross sections
of this reaction have been extracted from the total fission-like events by the
analysis of their mass and angular distributions.  The analysis of
these ``asymmetric shoulders'' in the mass-energy distributions
points to the quasifission nature of this component. The contribution
of the quasifission fragments with masses in the above-mentioned
range to the total mass distribution of fission fragments
increases, with respect to one of the symmetric compound
nucleus-fission, as the $^{48}$Ca projectile energy decreases.
In Fig. \ref{compcross}a we compare the experimental results for
the capture, quasifission and fusion-fission excitation functions
from Ref. \cite{Knyazheva} presented with the results of
calculations performed in the framework of the DNS model [see Refs.
\cite{Fazio04,Fazio05}. In this figure we present our results for
the fast-fission fragments, too. The contribution of the
fast-fission channel increases by increasing the bombarding energy
due to the increase in the angular momentum of the mononucleus.

The origination of the measured fission-like fragments at the
large bombarding energies is explained by the sum of the
quasifission (short dashed line), fusion-fission (dash-double
dotted line) and fast-fission (dash-dotted line) fragments.  At
lower energies the contribution of the fusion-fission to the yield
of binary fragments is small in comparison with the quasifission
contribution. The small calculated fusion-fission cross
section is explained by the large fission barrier ($B_{\rm
f}$=12.33 MeV) for the $^{202}$Pb nucleus according to the
rotating finite range model by A. J. Sierk \cite{Sierk} and by the
additional barrier $B_{\rm f}^{(\rm microscopic)}=-\delta
W=-(\delta W_{\rm saddle-point}-\delta W_{\rm gs})\cong 8.22$ MeV
caused by the nuclear shell structure. We conclude that the
experimental fusion-fission data obtained at low energy
collisions contain a huge contribution of
quasifission fragments with masses $A>83$ which show an isotropic
distribution as presented in Ref. \cite{Knyazheva}. This is not a
new phenomenon and it was discussed as a result of theoretical
studies, for example, in our previous papers
\cite{FazioMess,NasirovRauis}  and in Ref. \cite{AritomoNP744}.
 The experimental results confirming this  conclusion
appeared recently in Ref.  \cite{HindePRL101,KnyazhevaPNL}.
 At the large energy $E_{\rm c.m.}$=154 MeV ($E^*_{\rm CN}$=63 MeV) the
experimental values of the quasifission cross section are much
lower than that of the fusion-fission cross section.
A sufficient part of the quasifission fragments shows the behaviour
of the fusion-fission fragments: the mass distribution can reach
the mass symmetric region and their angular distribution can be
isotropic due to the possibility that the dinuclear system can
rotate by large angles for large values of its angular
momentum. The authors of Ref. \cite{Knyazheva} did not exclude
such a behaviour of the quasifission fragments. It is  difficult
to separate  the quasifission fragments from the fusion-fission
fragments when both, their mass and angle distributions, overlap in
the region of  symmetric masses.

It is well known that quasifission is the decay of the dinuclear
system into two fragments with symmetric or asymmetric masses. The
quasifission can take place at all values of the orbital angular
momentum leading to capture. Quasifission fragments formed at energies
above the Coulomb barrier  with a small angular momentum
contribute to the asymmetric part of the mass distribution.
Because the lifetime of the dinuclear system decreases by increasing
its excitation energy. The excitation energy is defined as
\begin{equation}
\label{Edns}
 E^*_{\rm DNS}(Z,A,\ell)=E_{\rm c.m.}-V_{\rm min}(Z,A,\ell),
\end{equation} where $V_{\rm min}(Z,A,\ell)$ is the minimum of the
potential well corresponding to the interaction of fragments with
the charge (mass) asymmetry $\{Z,Z_{\rm tot}-Z; A,A_{\rm
tot}-A\}$. As usually, the nucleus-nucleus potential
$V(Z,A,\ell,R)$ includes the Coulomb $V_{\rm Coul}(Z,A,R)$,
nuclear $V_{\rm N}(Z,A,R)$ and rotational $V_{\rm
rot}(Z,A,R,\ell)$ parts:
\begin{equation}
\label{Vpot} V(Z,A,\ell,R)=V_{\rm Coul}(Z,A,R)+V_{\rm
N}(Z,A,R)+V_{\rm rot}(Z,A,R,\ell),
\end{equation}
where $R$ is the distance between the centers of the nuclei. Details
of the calculation can be found in Refs. \cite{Fazio04,Nasirov05}.

 At low energies the projectile-like quasifission fragments with $A<70$ give a large
contribution to the cross section for the considered $^{48}$Ca+$^{154}$Sm
reaction  since the excitation energy of
the DNS is to small to shift the maximum of the mass distribution to
more mass symmetric configurations of dinuclear system. The
observed quasifission feature at low energies is connected with
the peculiarities of the shell structure of the interacting nuclei.
The increase in the beam energy leads to a decrease of the shell
effects and the yield of the quasifission fragments near the
asymmetric shoulders decreases. The main contribution to
quasifission moves to the symmetric mass distribution.
\begin{figure}
\vspace*{2.5cm}
\begin{center}
\resizebox{0.80\textwidth}{!}{\includegraphics{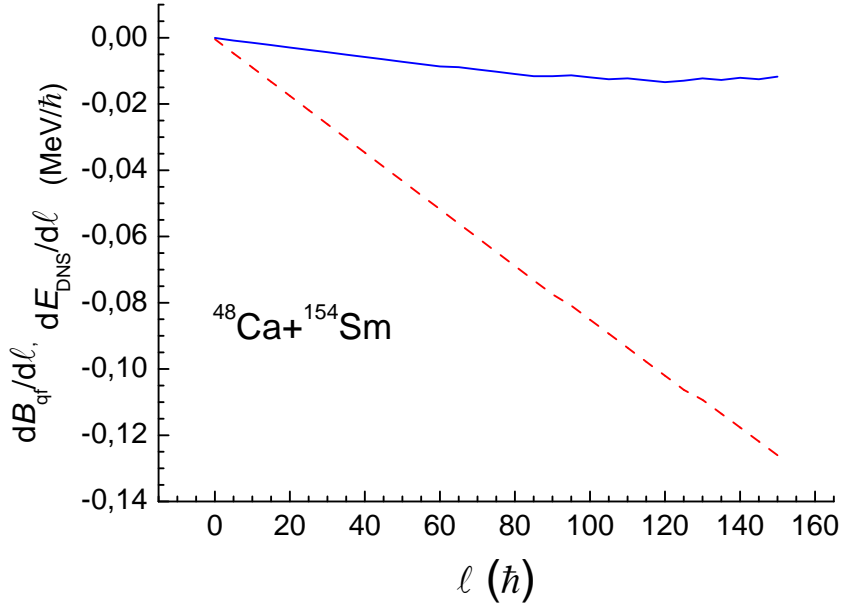}}
\vspace*{-5.0cm} \caption{\label{varEDNS-Bqf} The
decrease in the quasifission barrier $B_{\rm qf}(Z,\ell)$ (solid
line) and excitation energy $E^*_{\rm
DNS}(Z,\ell)$ (dashed line) of the dinuclear system  as a function of the
angular momentum for the $^{48}$Ca+$^{154}$Sm reaction.}
\end{center}
\end{figure}
A more interesting phenomenon at the same beam energies occurs for
the dinuclear system formed with large angular momenta. The
lifetime of the DNS can be long enough to reach large
rotational angles and to have a nearly isotropic angular distribution
of its quasifission fragments because $E^*_{\rm DNS}(Z,A,\ell)$
decreases as a function of angular momentum $\ell$, according to
its definition by formula (\ref{Edns}). Of course, the
quasifission barrier $B_{\rm qf}$ decreases by increasing  $\ell$
but it decreases  slower than $E^*_{\rm DNS}(Z,A,\ell)$ because we have
 \begin{equation}
 \label{inequal}
 |dE^*_{\rm DNS}(Z,\ell)/d\ell|>|dB_{\rm qf}/d\ell|,
 \end{equation}
 {\it i.e.} $E^*_{\rm DNS}(Z,\ell)$ decreases  faster than
 $B_{\rm qf}$ by increasing $\ell$ (see Fig. \ref{varEDNS-Bqf})
  at all beam energies.
This inequality follows from a comparison of the corresponding
derivations which can be obtained using Eqs.(\ref{Edns}) and
({\ref{Vpot}):
\begin{eqnarray}
\label{varEdns} \frac{dE^*_{\rm DNS}(Z,\ell)}{d\ell}&=&
-\frac{(2\ell+1)\hbar^2}{2(J_1+J_2+\mu R_{\rm m}^2)}.
\end{eqnarray}
Taking into account the definition of the quasifission barrier
$B_{\rm qf}(Z,\ell)=V_{\rm B}(Z,\ell)-V_{\rm min}(Z,\ell)$ which
yields
\begin{eqnarray}
\label{varBqf} \frac{dB_{\rm
qf}(Z,\ell)}{d\ell}&=&\frac{(2\ell+1)\hbar^2}{2(J_1+J_2+\mu
R_m^2)}-\frac{(2\ell+1)\hbar^2}{2(J_1+J_2+\mu R_{\rm B}^2)},
\end{eqnarray}
where $V_{\rm B}(Z,\ell)$ is the barrier of the nucleus-nucleus
potential which should be overcome at the decay of the dinuclear
system. $V_{\rm min}(Z,\ell)$ was discussed earlier;  $R_{\rm B}$
and $R_m$ are the positions of the barrier and minimum of the
potential well with $R_{\rm B} > R_{\rm m}$. From a comparison the right
sides of Eqs.(\ref{varEdns}) and ($\ref{varBqf}$) we obtain
Eq. \ref{inequal}. This inequality means that the dinuclear system
can rotate by large angles before it decays into two fragments if it is formed with
large angular momentum $\ell$. A large beam energy is needed to
form a dinuclear system with large values of $\ell$. The condition
is similar with the formation of super-deformed states of a
nucleus.
So, we can conclude that the quasifission fragments formed
at the decay of the fast rotating dinuclear system have nearly isotropic
angular distribution. If their mass distribution is in the region
of symmetric masses then the quasifission fragments are very
similar to the fusion-fission fragments and they are mixed with
the latter. This mechanism is responsible for the disappearance of
the ``asymmetric shoulders'' in the mass distribution of the fission
fragments from the $^{48}$Ca+$^{154}$Sm reactions at collision
energies $E_{\rm c.m.}> 154$ ($E^*_{\rm CN}> 63$ MeV).  The
experimental data, which were identified as fusion-fission
fragments by the authors of Ref. \cite{Knyazheva}, increase
strongly starting from the energies $E_{\rm c.m.}> 147$ ($E^*_{\rm
CN}> 57$) MeV. According to our results, a large part of
this increase belongs to the quasifission fragments (see Fig.
\ref{compcross}). So we stress that, in the $^{48}$Ca+$^{154}$Sm
reaction, the quasifission (short dashed line in
Fig.\ref{compcross}a) is the dominant channel in comparison with
the fusion-fission (dash-dotted line), total evaporation
residues (thick dotted line in Fig.\ref{compcross}b) and fast-fission
(dashed-double dotted line in Fig. \ref{compcross}a) channels. The
experimental data for the excitation function of the total
evaporation residues are taken from the paper by Stefanini {\it et
al.} \cite{Stefanini}. At energies $E_{\rm c.m.}<140$ MeV, our
capture cross section $\sigma_{\rm cap}$ overestimates the
experimental values of the capture cross section $\sigma^{(\rm
exp)}_{\rm cap}$ because the authors of Ref. \cite{Knyazheva}
excluded from their analysis the reaction products having mass
numbers outside the mass range $55<A<145$. Our studies
showed that capture events, {\it i.e.} events of the full momentum
transfer, can lead to yields of fragments with masses $A_{\rm
qf}<55$.  Consequently they lost a part of the capture cross
sections related to the contributions of the  quasifission
fragments with $A_{\rm qf}<55$. They determined
\begin{equation}
   \sigma^{(\rm exp)}_{\rm cap}(E_{\rm c.m.},A_{\rm qf})=
   \sigma^{(\rm exp)}_{\rm ER}(E_{\rm c.m.})+
   \sigma^{(\rm exp)}_{\rm f}(E_{\rm c.m.})+
   \sigma^{(\rm exp)}_{\rm qf}(E_{\rm c.m.},55<A_{\rm qf}<145)
\end{equation}
while the theoretical capture cross section includes the
contributions of all fragment yields, {\it i.e.} $4<A_{\rm
qf}<198$, from full momentum tranfer reactions
\begin{equation}
   \sigma_{\rm cap}(E_{\rm c.m.})=
   \sigma_{\rm ER}(E_{\rm c.m.})+
   \sigma_{\rm f}(E_{\rm c.m.})+
   \sigma_{\rm qf}(E_{\rm c.m.})+
   \sigma_{\rm fast-fission}(E_{\rm c.m.}).
\end{equation}
The experimental and theoretical capture cross sections come
closer when the beam energy increases due to three main
reasons: (i) a shift of the maximum of the DNS mass distribution to
the mass symmetric region: the amount of the lost part of the
capture cross section decreases; (ii) the quasifission fragments
within the $70<A_{\rm qf}<130$ range which are formed at the decay of
the DNS with large angular momentum show an isotropic angular
distributions being considered as fusion-fission fragments; (iii)
the fusion probability increases by increasing the beam energy due
to the inclusion of the contributions from collisions with
large orientation angles of the target-nucleus symmetry axis (see
Fig. \ref{PcnFig}) with respect to the beam direction. The
favourableness of the large orientation angles for the formation
of the compound nuclei was analyzed in Refs.
\cite{Nasirov05,Nasirov,Mandaglio}. This mechanism was earlier
suggested by \cite{Hinde95}. The probability of the compound
nucleus formation $P_{\rm CN}$ increases by increasing the
collision energy and excitation energy $E^*_{\rm CN}$, as seen in
Fig. \ref{PcnFig}. The presented results in Figs. \ref{compcross}
and \ref{PcnFig} are obtained by averaging over all orientation
angles of the symmetry axis of $^{154}$Sm which is a well deformed
nucleus ($\beta_2=0.341$). The role of the target orientation angle
relative to the beam direction during the formation of the
fusion-fission and ER products in the $^{48}$Ca+$^{154}$Sm
reaction was analyzed in Ref. \cite{Mandaglio}.
\begin{figure}
\vspace*{4.5cm}
\resizebox{1.1\textwidth}{!}{\includegraphics{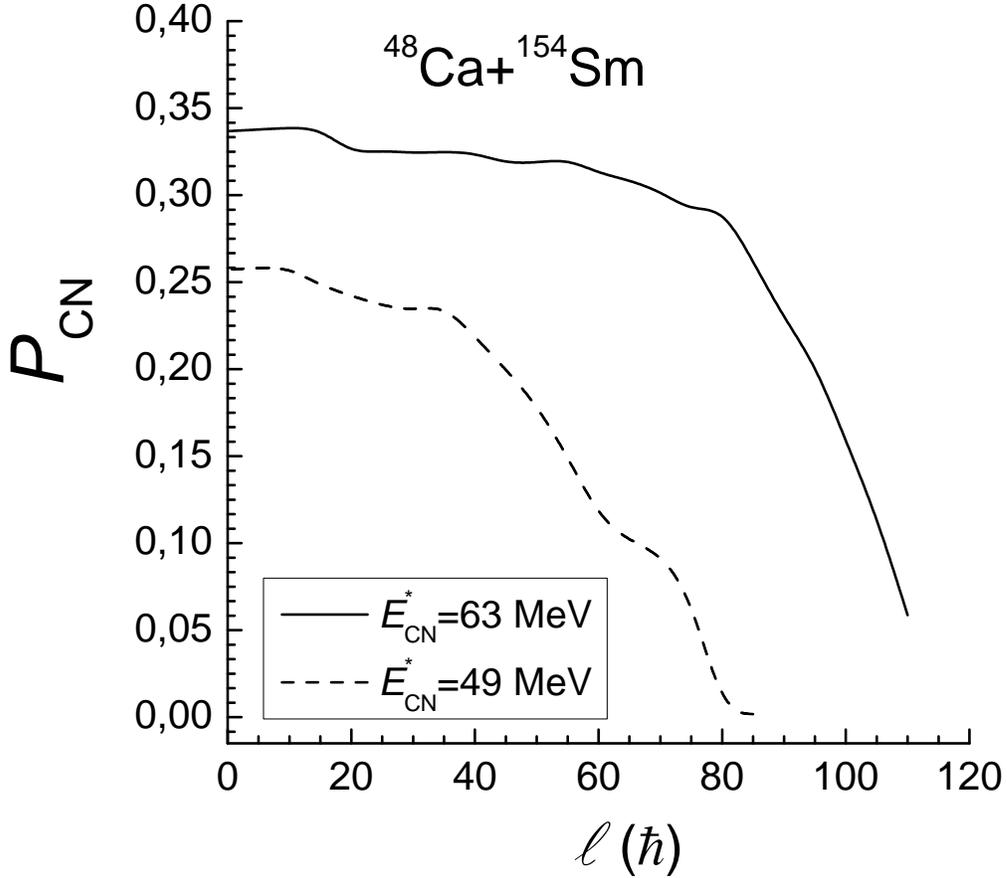}}
\vspace*{-6.3cm} \caption{\label{PcnFig} The probability $P_{CN}$
of the compound nucleus formation  as a function of the angular
momentum
 of dinuclear system $\ell$ at energies $E_{\rm c.m.}$ =138
and 154 MeV, corresponding to the excitation energies of the
compound nucleus $E^*_{\rm CN}$=49 and 63 MeV, respectively.}
\end{figure}
The decrease in $P_{\rm CN}$ by increasing the DNS angular
momentum $\ell$ is explained by the increase in the intrinsic
fusion barrier and decrease in the quasifission barrier by
increasing $\ell$ (see Refs. \cite{Fazio04,Fazio05}). So, we have
explained the large difference between the calculated and
experimental capture cross sections at low collision energies and the decrease in
this difference at
high collision energies. The experimental data of fission-like
fragments seem to include some part of the quasifission and
fast-fission fragments which overlaps with the mass and angular
distributions of the fusion-fission fragments.

\begin{figure}
\vspace*{3.0cm}
\resizebox{0.85\textwidth}{!}{\includegraphics{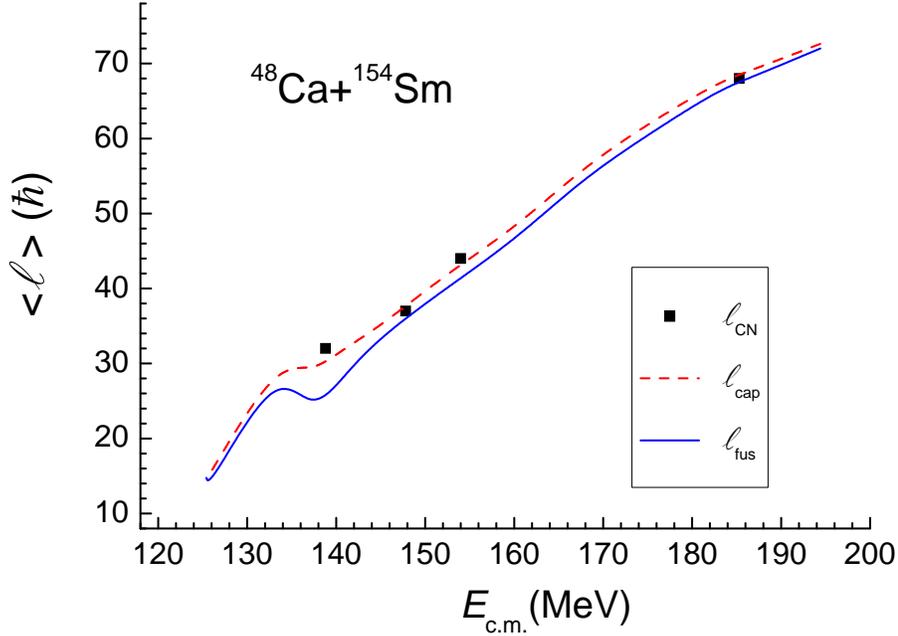}}
\vspace*{-5.5cm} \caption{\label{AverL} Comparison of the
calculated angular momentum distribution of the compound nucleus $^{202}$Pb
formed in the $^{48}$Ca+$^{154}$Sm reaction
with the experimental data from Ref.\cite{Knyazheva}. The presence
of the quasifission contribution in the measured data is
noticeable at low energies.}
\end{figure}

The agreement of the results for the  angular momentum
distributions with the measured ones in Ref. \cite{Knyazheva}
confirms that the angular momentum distributions of the compound
nuclei obtained by us is correct. The results of this comparison are
presented in Fig. \ref{AverL}. The deviation of the results for
$\ell_{\rm CN}$ of this work from the experimental data at $E_{\rm
c.m.}$=138 MeV is explained by large contributions of
quasifission fragments.

\subsection{About missing quasifission events in the $^{48}$Ca+$^{144}$Sm reaction.}

The authors of Ref. \cite{Knyazheva} concluded from the study of mass-angle
distributions in $^{48}$Ca+$^{144}$Sm reactions that
there are not quasifission contributions to  the mass distribution
in the analyzed range $60 < A < 130$. The theoretical calculations
in this work show that quasifission occurs in this reaction
causing the hindrance for the formation of the compound nucleus. But this
hindrance is less active than the one in the case of the reaction with
$^{154}$Sm. The presence of the quasifission feature is expected
from the non zero value of the intrinsic fusion barrier
$B^*_{fus}$ which is found from the driving potential calculated
for the $^{48}$Ca+$^{144}$Sm reaction. The results for the
capture, complete fusion, evaporation residue, fusion-fission and
fast-fission cross sections are presented in Fig.
\ref{CrossSec144Sm}.
\begin{figure}
\vspace*{3.5cm}
\begin{center}
\resizebox{0.90\textwidth}{!}{\includegraphics{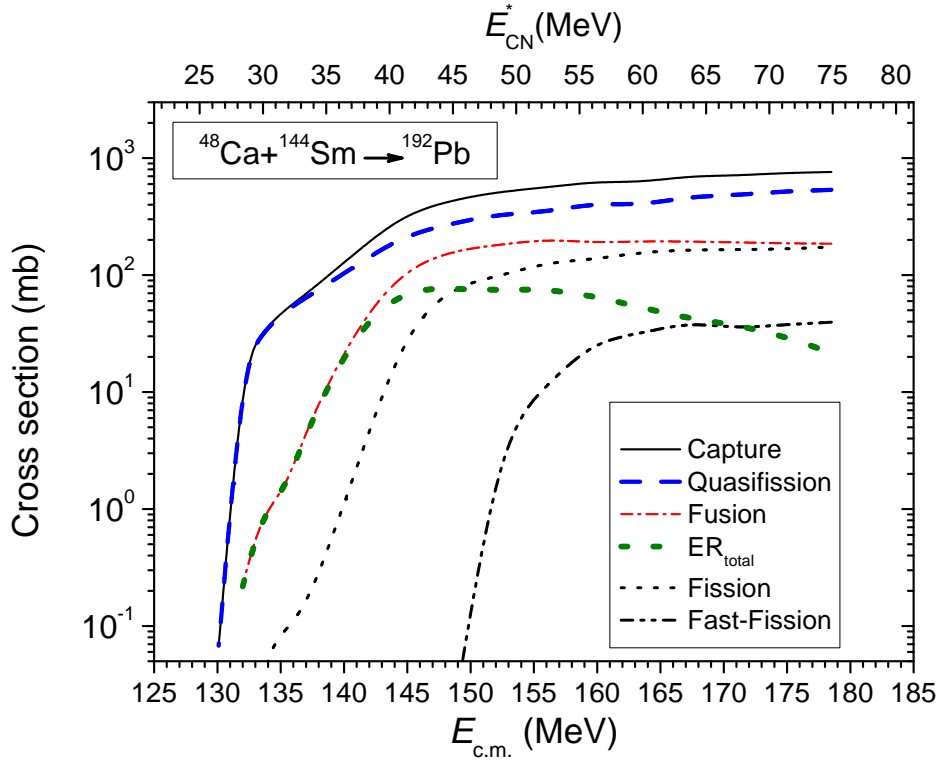}}
\vspace*{-5.0cm} \caption{\label{CrossSec144Sm} The results by the
DNS model for the capture, complete fusion, quasifission, fission,
evaporation residue, and fast-fission cross sections for the
$^{48}$Ca+$^{144}$Sm reaction.}
\end{center}
\end{figure}
It is seen that the theoretical results indicate  large
contributions of quasifission to the capture cross section.
Unfortunately, the authors of Ref. \cite{Knyazheva} did not
investigate the fusion or
evaporation residue cross sections which could be compared with our
results. The contradiction between our results and the conclusion
of the authors of Ref. \cite{Knyazheva} from their analysis of the
selected experimental data about a presence or lack of quasifission in the
$^{48}$Ca+$^{144}$Sm reaction may be removed if we answer the
question why the quasifission events were not observed? There are
two reasons: (i)  one part of the mass distribution of the
quasifission fragments is outside the analyzed range of $60 < A <
130$; (ii) another part of the quasifission fragments is mixed
with the fusion-fission fragments and has similar isotropic
distributions. The masses of the missing quasifission fragments are
in the mass range $48 < A < 60$. This range is outside the
analyzed range and, therefore, the missing fragments can not show
the presence of quasifission. The isotope $^{144}$Sm is a magic
nucleus with neutron number $N$=82. Therefore, the concentration
of the asymmetric mode of the quasifission fragments in the mass
range $48 < A < 60$ is explained by the effect of the shell
structure of the double magic projectile-nucleus $^{48}$Ca and
magic target-nucleus $^{144}$Sm on the mass distribution of the
reaction fragments. As a result the mass distributions of the
products of deep-inelastic collisions and asymmetric quasifission
overlap in this mass range.

This case is similar to the $^{48}$Ca+$^{208}$Pb reaction where
the presence of the quasifission feature was doubtful (see Ref.
\cite{faziolett} and references therein). But our investigation showed
that due to the collision of the double magic
 $^{48}$Ca and $^{208}$Pb nuclei the mass distribution of the
quasifission fragments is concentrated around the initial masses
\cite{faziolett} because the potential energy surface has a local
minimum in this region. Moreover the products of these processes
have similar  angular distributions for collisions with small
values of the orbital angular momentum $\ell$ but they can be
separated by the total kinetic energy distributions. In the
quasifission process the full momentum transfer takes place. In
collisions with large angular momentum the angular
distributions of the products of the quasifission and
deep-inelastic collisions should be different due to the long
lifetime of the dinuclear system formed at the capture stage of reaction.
This kind of studies will be useful to investigate the mechanism of the full
momentum transfer reactions.

 Concerning the second reason, the calculation of the mass
distribution of quasifission fragments for the
$^{48}$Ca+$^{144}$Sm reaction showed that there is another group
of fragments which is placed in the mass-symmetric region and
is mixed with the fusion-fission fragments.

We suggest  to measure the cross section of the evaporation
residues and compare with the corresponding data obtained in Ref.
\cite{Stefanini} for the $^{48}$Ca+$^{154}$Sm reaction. We expect
that the excitation function of evaporation residues of the latter
reaction will be higher than the one which could be obtained for the
$^{48}$Ca+$^{144}$Sm reaction. This will be the evidence for the
presence of quasifission fragments in $^{48}$Ca+$^{144}$Sm
reaction. Of course, the fact that the compound nucleus $^{192}$Pb
formed in the last reaction has a smaller number of neutrons leads to
a decrease in the evaporation residue cross sections but the effect
of quasifission should be stronger than the effect of the
difference in the neutron numbers in the compound nuclei $^{192}$Pb
and $^{202}$Pb. In Fig. \ref{CrossSec144Sm} we present our
theoretical results  for the excitation function of the
evaporation residues (thick short-dashed line) of the
$^{48}$Ca+$^{144}$Sm reaction. A comparison of the results of
$\sigma_{ER}$ for the reactions with $^{154}$Sm (see
Fig.\ref{compcross}) and $^{144}$Sm (Fig. \ref{CrossSec144Sm})
shows that the values of the former reaction are larger than the ones
of the latter reaction. The fusion cross sections are nearly the
same but the capture excitation function for the reaction with
$^{144}$Sm is lower than the one for the
$^{48}$Ca+$^{154}$Sm reaction because the attractive nuclear
forces are stronger in the more neutron rich system. So we can
conclude that according to our theoretical studies there are
quasifission events in the $^{48}$Ca+$^{144}$Sm reaction.
The authors of Ref. \cite{Knyazheva} did not observe them because the
mass distribution of the first group of quasifission fragments
was outside the mass range $60 < A < 130$. The second group of the
quasifission fragments has an overlap in the mass-angle distributions
with the fusion-fission fragments in the studied mass range.

\subsection{About a lack of quasifission in the $^{16}$O+$^{186}$W reaction.}

To check the reliability of our calculation method we analyzed
also the $^{16}$O+$^{186}$W reaction where the complete fusion is
the main channel among capture reactions. Indeed, the driving
potential of this reaction does not show an intrinsic fusion
barrier $\ell=0$ excluding a small barrier connected with the effect of the
odd-even nucleon numbers. But an intrinsic fusion barrier for
 can arise
at large values of the orbital angular momentum $\ell$. The
results of the calculation for the capture, complete fusion,
evaporation residue and fusion-fission cross sections are
presented in Fig. \ref{CrossSecO186W}. One can see that up to
$E_{\rm c.m.}=90$ MeV the excitation functions of capture and
complete fusion are mainly the same because  the contribution of
quasifission is very small (more than one order lower). Certainly,
the evaporation residue cross section is  enough large and it
decreases at large values of the beam energy due to the decrease in
the stability of the heated and rotating compound nucleus. At
$E_{\rm c.m.}> 83$ MeV the fission cross section is higher than
the ER cross section. The fast-fission contribution is small and it appears
appreciably at $E_{\rm c.m.}> 100$ MeV.  The mass distribution of
quasifission fragments does not reach the mass symmetric region and
consequently there is not an overlap with fusion-fission fragments.
The measured fission fragments correspond to a pure fusion-fission
channel.
\begin{figure}
\vspace*{3.5cm}
\begin{center}
\resizebox{0.90\textwidth}{!}{\includegraphics{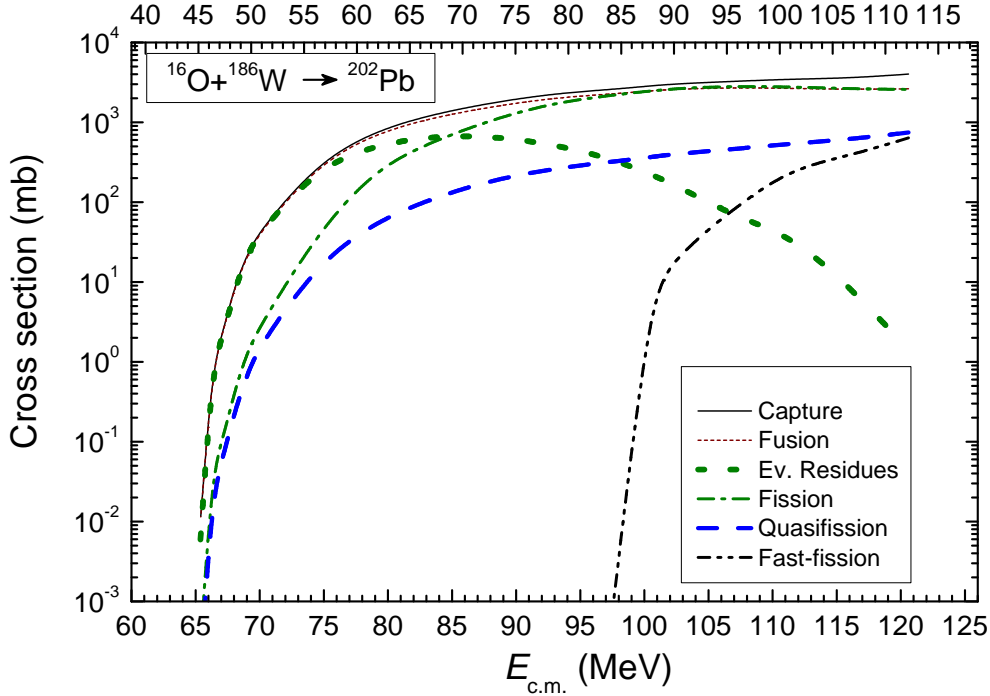}}
\vspace*{-5.0cm} \caption{\label{CrossSecO186W} The results of the
DNS model for the capture, complete fusion, quasifission, fission,
 evaporation residue, and fast-fission cross sections for the
$^{16}$O+$^{186}$W reaction.}
\end{center}
\end{figure}

Concluding this section, in Fig. \ref{Pcn3reac} we present the calculated fusion
probability $P_{\rm CN}$ for the discussed reactions as a function
of the collision energy relative to the corresponding interaction
barrier for each reaction in Fig \label{Pcn3reac}. It is seen that in the mass asymmetric
$^{16}$O+$^{186}$W reaction the fusion probability is large.
The mass distributions of quasifission fragments are near the target
and projectile masses and they are not mixed with the
fusion-fission fragments. Therefore, all fission fragments near
the mass symmetric region belong to the fission of the compound nucleus.
The quasifission process evidently takes place in the reactions
$^{48}$Ca+$^{144}$Sm and $^{48}$Ca+$^{154}$Sm. It is
more intense in the latter reaction. The lack of quasifission
events in the experimental studies of the $^{48}$Ca+$^{144}$Sm
reaction or disappearance of quasifission events by increasing the
beam energy are connected with the measurement
and analysis of the experimental data. More advanced
experimental methods can be developed in order to study the
quasifission feature in the case that the mass-angle distributions of the
quasifission and fusion-fission fragments strongly overlap in
the mass symmetric region.

\begin{figure}
\vspace*{3.5cm}
\begin{center}
\resizebox{0.90\textwidth}{!}{\includegraphics{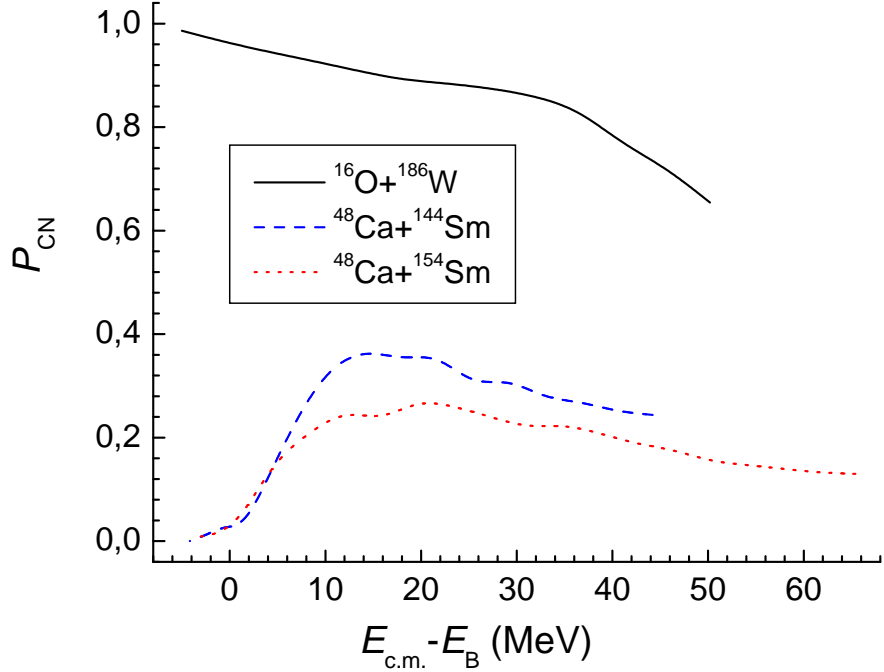}}
\vspace*{-5.5cm} \caption{\label{Pcn3reac} The DNS model results
for the fusion probability  $P_{CN}$ for the $^{16}$O+$^{186}$W,
$^{48}$Ca+$^{144}$Sm  and $^{48}$Ca+$^{154}$Sm reactions as a
function of the collision energy relative to the interaction
barriers corresponding to each of reactions.}
\end{center}
\end{figure}

\section{Role of the charge asymmetry and nuclear shell structure
in the yields of reaction products}

The theoretical method based on the dinuclear system concept is used
to analyze capture, complete fusion, quasifission and fast-fission
contributions in the reactions with massive nuclei and can be applied to
estimate and make predictions which of the reactions $^{54}$Cr+$^{248}$Cm,
$^{58}$Fe+$^{244}$Pu, and $^{64}$Ni+$^{238}$U is most
preferable to synthesize the superheavy element $Z$=120.

The advantage of the cold fusion reactions is a large survival
probability in the emission of one or two neutrons from the weak heated CN.
This way was used to obtain the first superheavy elements $Z$=110
(darmstadtium), 111 (roentgenium), 112 (see Ref.
\cite{SHofGMun,Morita04}), and $Z$=113 \cite{Morita07}. The grave
disadvantage of "cold fusion" reactions is the dominance of the
quasifission process as channel causing a hindrance in transforming
the DNS into a compound nucleus. According to the DNS model, for
the more mass symmetric reactions, the intrinsic fusion barrier is
larger in comparison to the one for  mass asymmetric reactions
\cite{Fazio04,Fazio05,Giardina00}. But the hindrance  caused by
quasifission is not so strong in mass asymmetric
"hot fusion" reactions. This is supported by the synthesis of the
even heavier new elements $Z$=114, 115, 116, 118 which were observed in
reactions with $^{48}$Ca ion-beams on
$^{244}$Pu, $^{243}$Am, $^{245}$Cm and $^{249}$Cf
actinide targets at the Flerov Laboratory of Nuclear Reactions of JINR
in Dubna \cite{Oganessian04,Oganessian07}. The cross section for the
synthesis  of the new element 118 was about 0.5 pb in the
 $^{48}$Ca+$^{249}$Cf reaction \cite{Oganessian07}. The results of the
 calculations for the cross sections of formation of the dinuclear system,
 compound nucleus and evaporation residues in this reaction are
 presented in Fig. \ref{ER118}. The relatively good agreement between the
 experimental data and our estimations for the evaporation residues
 gives hope to use the method based on the dinuclear system
 concept to investigate which of the reactions
$^{54}$Cr+$^{248}$Cm, $^{58}$Fe+$^{244}$Pu, and
$^{64}$Ni+$^{238}$U is preferable to synthesize the
superheavy element $Z$=120.

\begin{figure}
\vspace*{-1.5cm}
\resizebox{0.9\textwidth}{!}{\includegraphics{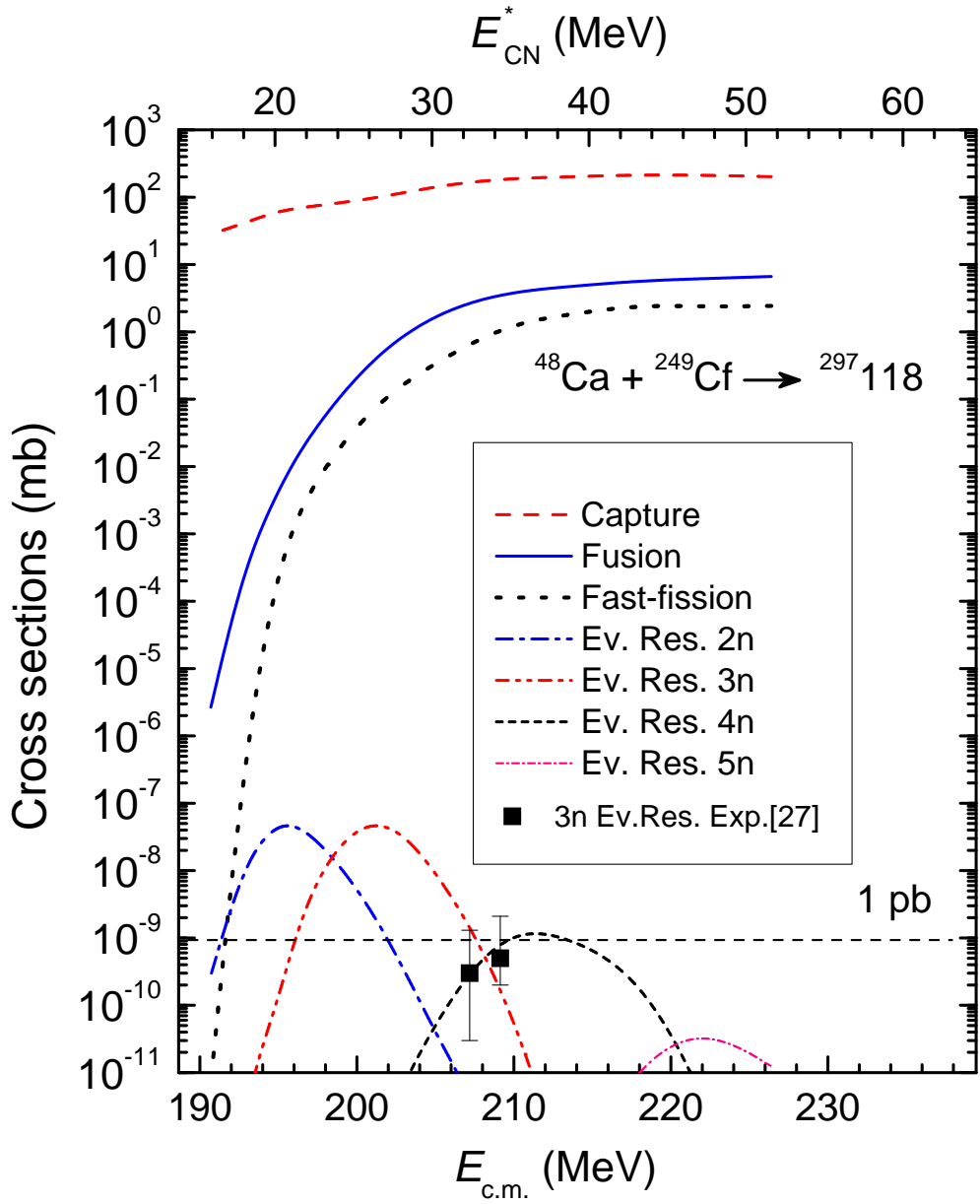}}
\vspace*{-5.0cm} \caption{\label{ER118} Excitation functions of the
formation of the dinuclear system (capture), compound nucleus (fusion)
and fast-fission fragments  in the $^{48}$Ca+$^{249}$Cf reaction
(upper part of the figure). Comparison of the calculated evaporation
residues in this reaction with the experimental data from Ref.
\cite{Oganessian07} (lower part of the figure)}.
\end{figure}

 In Ref.\cite{Fazio05}, we discussed the difference in the yields of
evaporation residues in different reactions leading to the same
compound nucleus. It was shown that the relationship between the
excitation energy $E^*_{\rm DNS}$  and intrinsic barrier $B^*_{\rm
fus}$ of the dinuclear system indicates which reaction is better
to produce an evaporation residue with a large cross section.
The results of the
calculations in this work show that among the three reactions
$^{54}$Cr+$^{248}$Cm, $^{58}$Fe+$^{244}$Pu, and
$^{64}$Ni+$^{238}$U the first one is preferable for the
synthesise the superheavy element $Z$=120 in comparison with
the two last ones.

\begin{figure}
\vspace*{-0.5cm}
\resizebox{0.8\textwidth}{!}{\includegraphics{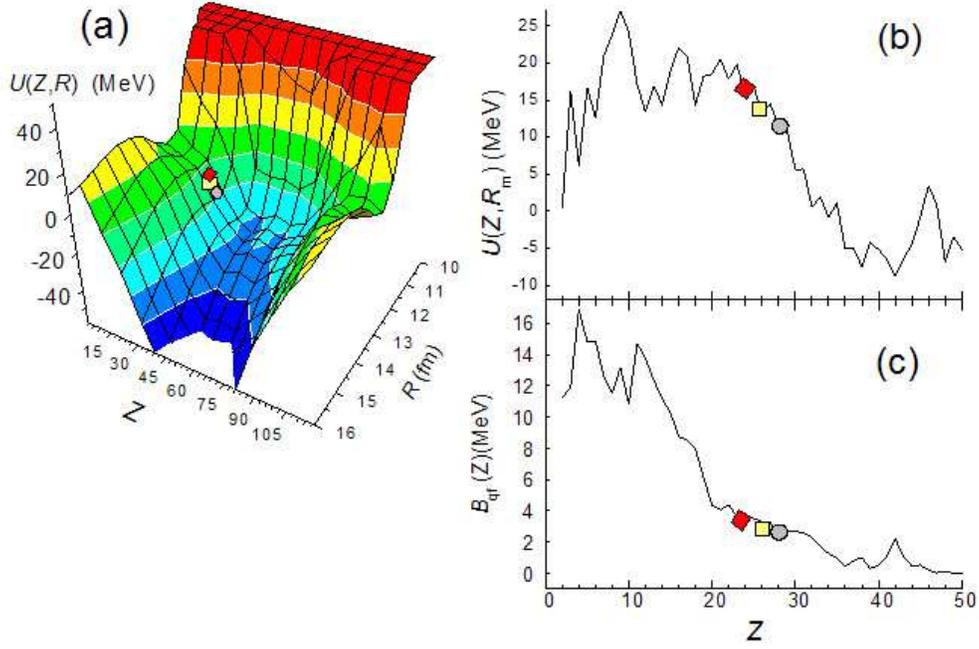}}
\vspace*{-0.2cm} \caption{\label{PES120} Potential energy surface
$U(Z,R)$ calculated for the DNS configurations leading to the
formation of the compound nucleus $Z$=120 and $A$=302 as a
function of the fragment charge number $Z$ and  relative distance
between the centers of fragments $R$ (a), driving potential
$U(Z,R_m)$ (b) and quasifission barriers $B_{\rm qf}$ as a
function of $Z$ (c). The initial points for the dinuclear systems
formed in the $^{54}$Cr+$^{248}$Cm, $^{58}$Fe+$^{244}$Pu, and
$^{64}$Ni+$^{238}$U reactions are shown by a diamond, a rectangle and
a circle, respectively.}
\end{figure}

 The analysis of the reactions with massive nuclei show that
the mass asymmetry, shell structure and orientation angles of the
symmetry axes of the initial colliding nuclei play a crucial role
in the formation of reaction products at the final stage of the
process \cite{Fazio04,Fazio05,Nasirov,Giardina00,Mandaglio}. The
failure in the synthesis of the superheavy element $Z=118$ in the ``cold
fusion" reaction $^{86}$Kr+$^{208}$Pb at the Lawrence Berkeley
Laboratory is explained by the very large value of $B^*_{fus}$
 of the dinuclear system consisting of the $^{86}$Kr and
$^{208}$Pb nuclei. In other words,  the
touching point is far from the saddle-point corresponding
to the compound nucleus $^{294}$118. Due to the small quasifission
barrier the lifetime of the  dinuclear system is short and
its excitation energy is not sufficient to reach the saddle-point. An
increase of the beam energy does not supply the needed
excitation energy because by increasing the beam energy the
capture events are lost due to the smallness of the potential well in the
nucleus-nucleus interaction between the nuclei. The calculated
friction coefficient is not sufficient to trap the projectile
 into the small potential well  at large energies.
Details of this phenomenon are explained, for example, in Fig. 1 of
Refs. \cite{Fazio05} and \cite{Giardina00} or Fig.2 in Ref.
\cite{Nasirov}.  The dynamics of the entrance channel was discussed in
the last cited papers. This circumstance proves the importance of
a correct calculation of the potential energy surface and the
friction coefficients. Their quantities determine the distributions
of the angular momentum and excitation energy  between the
fragments forming the DNS.

In the DNS model the capture and fusion stages are studied in
detail to analyze experimental data. The observed hindrance to
complete fusion in reactions with massive nuclei is connected with
the intrinsic fusion barrier $B^*_{\rm fus}$  which is sensitive
to the mass asymmetry and shell structure of the nuclei in the
entrance channel \cite{Fazio04,Fazio05}. The fusion barrier is
determined by the peculiarities of the potential energy surface
$U(Z,A,R)$  \cite{Fazio04,Fazio05}  calculated  for the
DNS leading to $Z$=120 and $A$=302. The potential energy surface
is a sum of the mass balance for DNS fragments and
the nucleus-nucleus interaction potential $V(Z,R)$:
\begin{equation}
U(Z,A,R)=B_1(Z)+ B_2(Z_{\rm tot}-Z)+V(Z,R)-B_{\rm CN}(Z_{\rm
tot}),
\end{equation}
where $B_1(Z)$,  $B_2(Z_{\rm tot}-Z)$ and $B_{\rm CN}(Z_{\rm
tot})$ are the ground state binding energies of the DNS fragments
1 and 2, and compound nucleus, respectively
\cite{Fazio04,Fazio05}. The potential energy surface, driving
potential and  quasifission barriers for the reactions leading to
the CN $^{302}$120 are presented in Fig. \ref{PES120}. The
characteristics of the entrance channel as mass (charge)
asymmetry,  shell structure and the orientation angles of the symmetry
axis (for the deformed nucleus) of colliding nuclei  are
important for the formation probability and angular momentum
distribution of the compound nucleus. The survival
probability of the heated and rotating compound nucleus depends on
its angular momentum $\ell_{\rm CN}$ and excitation energy
$E^*_{\rm CN}$ \cite{Fazio04,Fazio05}. The small intrinsic fusion
barrier $B^*_{\rm fus}$ and large quasifission barrier $B_{\rm
qf}$  lead to a large fusion probability. Fig.
\ref{PES120} shows that the conditions are satisfied better for the
$^{54}$Cr+$^{248}$Cm reaction.

\begin{figure}
\vspace*{-1.0 cm}
\resizebox{0.65\textwidth}{!}{\includegraphics{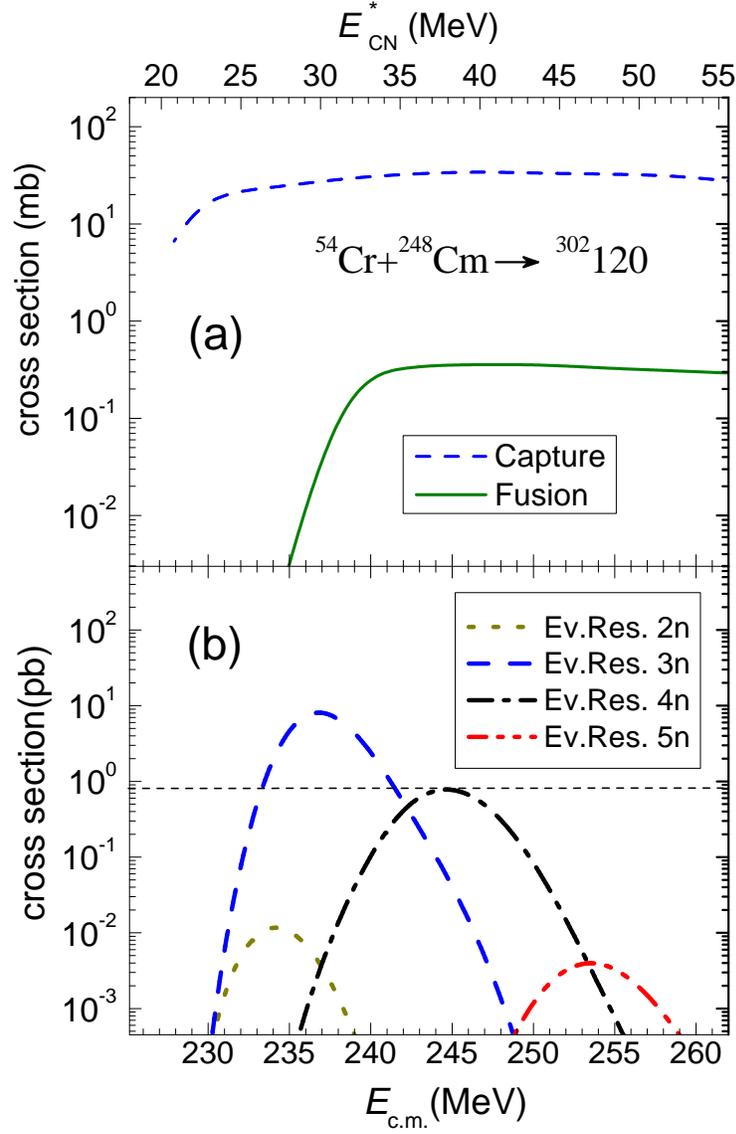}}
\vspace*{-0.4 cm} \caption{\label{fig6} Calculated excitation functions
for the capture and fusion  (a) and for the formation of the
evaporation residues in the 2n, 3n, 4n, and 5n channels (b) in the
the $^{54}$Cr+$^{248}$Cm reaction.}
\end{figure}

It is seen from these figures that the $^{54}$Cr+$^{248}$Cm
reaction is advantageous due to the small intrinsic fusion barrier
$B^*_{\rm fus}$ because it is placed close to the maximum
("saddle-point" to fusion) on the way in the fusion valley (to
reach small values of $Z$) (Fig. \ref{PES120}a). The quasifission barrier for this
reaction is larger because it is more  asymmetric in the charge (mass)
than in the two other reactions (see Fig.\ref{PES120} c).

 The results of the calculations of capture, complete fusion and evaporation residue
formation for the reactions $^{54}$Cr+$^{248}$Cm, $^{58}$Fe+$^{244}$Pu and
$^{64}$Ni+$^{238}$U are presented in Figs. \ref{fig6},
\ref{fig7} and \ref{fig8}, respectively. We stress that the
deformed shape of the initial projectile-target nuclei and the
possibility of collisions with different orientation angles of
their symmetry axes relative to the beam direction are taken into account as
in Ref. \cite{Nasirov}. The comparison of the Figs. \ref{fig6}, \ref{fig7}
and \ref{fig8} shows that the $^{54}$Cr+$^{248}$Cm reaction is
more favorable to synthesize the new superheavy element $Z$=120
because the predicted excitation functions of the 2n and 3n
evaporation residue channels are much larger than the maximal
values for the other two  reactions. We may state that the
estimated evaporation residue cross sections are in the range of
possibility of the detection systems used in the Flerov Laboratory
of nuclear reactions of JINR (Dubna, Russia), SHIP of GSI
(Darmstadt, Germany) and RIKEN (Japan).
\begin{figure}
\vspace*{-1.0 cm}
\resizebox{0.65\textwidth}{!}{\includegraphics{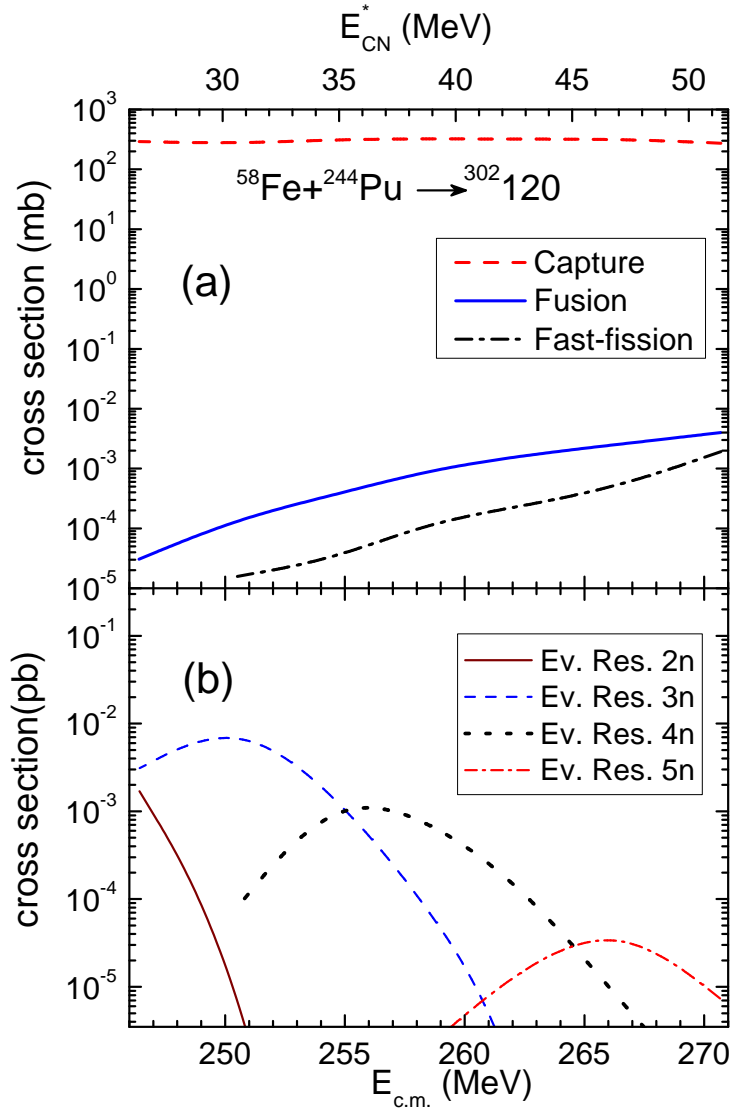}}
\vspace*{-0.4 cm} \caption{\label{fig7} Calculated excitation functions for the
capture, fusion and fast-fission   (a) and for the
formation of the evaporation  residues in the 2n, 3n, 4n, and 5n
channels (b) in  the $^{58}$Fe+$^{244}$Pu reaction.}
\end{figure}
\begin{figure}
\vspace*{-1.0 cm}
\resizebox{0.65\textwidth}{!}{\includegraphics{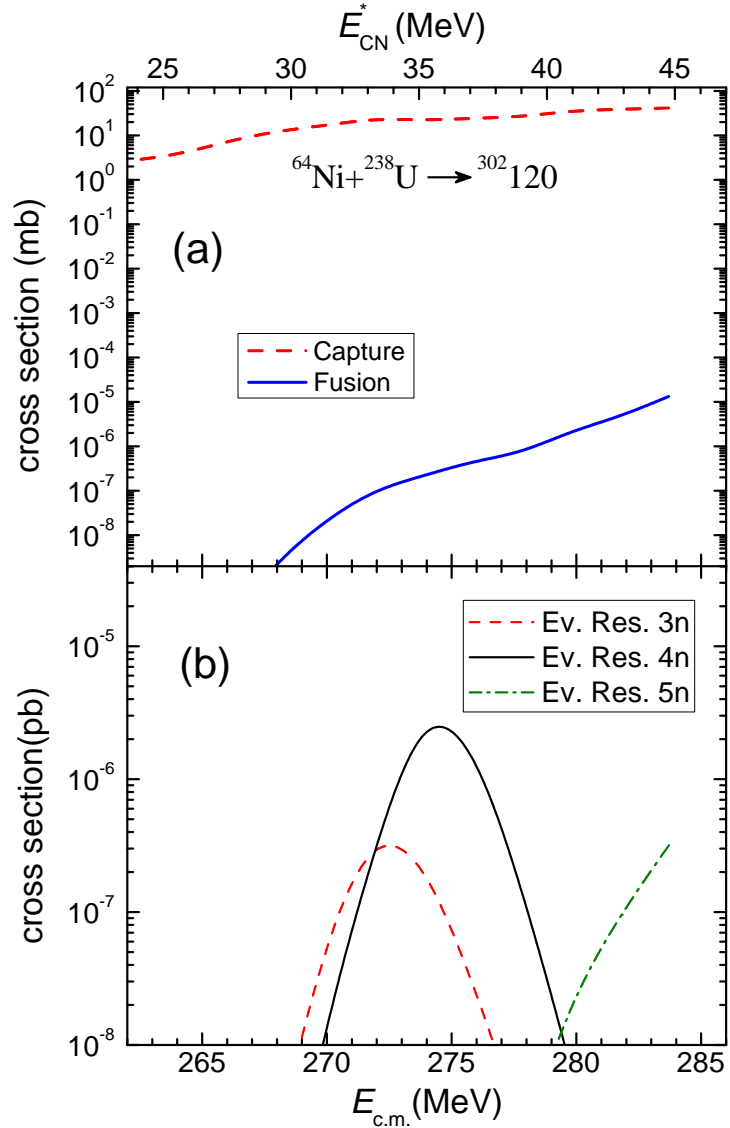}}
\vspace*{-0.4 cm} \caption{\label{fig8} Calculated excitation functions
for the capture and fusion  (a) and for the formation of the
evaporation residues in the 3n, 4n, and 5n channels (b) in
the  $^{64}$Ni+$^{238}$U reaction.}
\end{figure}

\section{Conclusions}

In this paper we have analyzed the reasons for the missing  of the
quasifission features in the $^{48}$Ca+$^{144}$Sm reaction and the
disappearance of the quasifission features in the
$^{48}$Ca+$^{154}$Sm reaction at  collision energies increasing
from $E_{\rm c.m.}=154$ MeV to larger values in the
experiments investigated in Ref. \cite{Knyazheva}. Our studies and
analysis of complete fusion  and formation of evaporation residues
showed the presence of quasifission in both of these reactions.
The experimental results for the capture, quasifission and
fusion-fission excitation functions from Ref. \cite{Knyazheva} and
data on the evaporation residues for this reaction from Ref.
\cite{Stefanini} were compared with the results of calculations
performed in the framework of the DNS model (see Refs.
\cite{Fazio04,Fazio05}).  The appearance of the measured
fission-like fragments at large bombarding energies is
explained by the sum of the quasifission, fusion-fission  and
fast-fission  fragments.  We conclude that the experimental
fusion-fission data obtained at low collision energies contain
a huge amount of contributions of quasifission fragments with
masses $A>83$ which show isotropic angular distributions as
presented in Ref. \cite{Knyazheva}. The smallness of the
calculated fusion-fission cross section is explained by the large
fission barrier for the $^{202}$Pb nucleus, $B_{\rm f}$=12.33 MeV,
according to the rotating finite range model by A. J. Sierk
\cite{Sierk} and the additional barrier $B_{\rm f}^{(\rm
microscopic)}\cong 8.22$ MeV caused by the nuclear shell
structure.  The quasifission fragments formed in the decay of the fast
rotating dinuclear system have near isotropic angular
distribution. Such fragments are mixed with the fusion-fission
fragments if the mass distributions of both processes  overlap in
the region of symmetric masses. This mechanism is responsible for
the disappearance of the ``asymmetric shoulders'' in the mass
distribution of the fission fragments of the $^{48}$Ca+$^{154}$Sm
reactions at collision energies $E_{\rm c.m.}> 154$ ($E^*_{\rm
CN}> 63$ MeV).  The experimental data, which were identified as
fusion-fission fragments by the authors of Ref. \cite{Knyazheva},
increase strongly starting from the energies $E_{\rm c.m.}> 147$
($E^*_{\rm CN}> 57$) MeV. According to our results, a sufficient
part of this increase belongs to the quasifission fragments (see
Fig. \ref{compcross}). The calculated excitation function of the
evaporation residues is in good agreement with the available
experimental data from  Ref. \cite{Stefanini}. Its values for
large collision energies decrease strongly due to the decrease in
the fission barrier of the heated and rotating compound nucleus if
 its excitation energy and angular momentum increases.

 The contradiction between our results
and conclusions of the authors of Ref. \cite{Knyazheva} from the
analysis of the selected experimental data about the lack of the
quasifission process in the $^{48}$Ca+$^{144}$Sm reaction is
connected with two main reasons: (i) the quasifission
fragments have a mass distribution with the maximum outside of
the analyzed range $60 < A < 130$; (ii) the quasifission fragments
are mixed with the fusion-fission fragments and have similar
isotropic distributions. The concentration of the first group of
quasifission fragments in the mass range $48 < A < 60$ is
explained by the effect of the shell structure of the double magic
projectile-nucleus $^{48}$Ca and magic target-nucleus $^{144}$Sm
on the mass distribution of the reaction fragments. Therefore,
the mass distributions of the products from deep-inelastic
collisions and quasifission overlap in this mass range. A similar
case was analyzed for the $^{48}$Ca+$^{208}$Pb reaction in
Ref.\cite{faziolett}.  Products of the decay of the
long-lived dinuclear system which were formed at large values of
the  angular momentum contributing to the  mass range $60 < A <
130$ seemed to be considered as the products of the fusion-fission
reactions because the products of both processes have overlap
of the mass and angular distributions.

The results obtained for the $^{16}$O+$^{186}$W reaction show
that a hindrance for complete fusion \cite{Knyazheva} is
negligible. Using the experience obtained in the analysis of the
above-mentioned reactions we estimate the most preferable reaction for
the synthesis of the superheavy element $Z=120$.
Among the three studied reactions,
$^{54}$Cr+$^{248}$Cm, $^{58}$Fe+$^{244}$Pu, and
$^{64}$Ni+$^{238}$U, the first one is most preferable for the
synthesis of the element $Z$=120. Because a more asymmetric
reaction it has a smaller intrinsic fusion barrier and
a larger quasifission barrier. These lead to a larger fusion cross section.
The expected cross section for the synthesis of superheavy element
$Z$=120 in the $^{54}$Cr+$^{248}$Cm reaction is more than 1 pb for
the 2n and 3n evaporation channels at $E_{\rm c.m.}$=233--245 MeV.

{\bf Acknowledgements}

 This work was performed partially under the financial support of
 the DFG, RFBR  and INTAS which are thanked very much by the author
 AKN.  Authors AKN and GG are also grateful to the Fondazione Bonino-Pulejo of
 Messina for the support received in the collaboration between the Dubna and Messina groups.
 Author AKN is grateful to the SHIP group in GSI and Physics Department of
 the Messina University for warm hospitality during his staying.

\end{document}